# The pandemic of viruses with a long incubation phase in the small world

P.A. Golovinski[1]

Voronezh State Technical University
394006 Voronezh, 20-letiya Oktyabrya Street, 84

A model of the spread of viruses in selected city and in a network of cities is considered, taking into account the delay caused by the long incubation period of the virus. The effect of delay effects is shown in comparison with pandemics without such delay. A temporary asymmetry of the spread of infection has been identified, which means that the time for a pandemic to develop significantly exceeds the time for its completion. Model calculations of the spread of viruses in a network of interconnected large and small cities were carried out, and dynamics features were revealed in comparison with the spread of viruses in a single city, including the possibility of reinfection of megalopolis.

## 1. Introduction

The spread of viruses in systems with a network structure and the emergence of pandemics is not only of fundamental scientific interest, but is a vital problem in natural biological populations and for the human community, as well as in artificial computer systems. The danger of global pandemics gives particular urgency to the problem, and the COVID-19 coronavirus pandemic became a clear manifestation of this. The direct experiment in such complex systems is practically impossible, therefore, in addition to studying and analyzing specific cases, there remains the way of a rather limited laboratory and alternative mathematical modeling of pandemics. Due to the safety and speed of assessing the development of situations, one of the most promising tools for studying the emergence, growth, spread and extinction of pandemics is their mathematical and computer modeling. The simplest model of the spread of the virus refers to a homogeneous system in which the spread rate depends on the number of infected carriers and the number of individuals not yet infected. The spread of viruses in such a system is wave-like and is described by the Fisher-Kolmogorov-Petrovsky-Piskunov equation [1-4]. However, real natural, socio-economic and artificial technical systems are highly spatially heterogeneous, which greatly complicates the analysis and understanding of the processes of spread of viruses in them [5, 6]. One of the most important factors influencing the understanding and control of pandemics is the time delay of the process, when a certain time elapses from the moment of infection to the moment of active manifestation of the virus, after which the individual becomes a source of infection [7, 8]. The effects of delay mask the cause-effect relationships in the dynamics of the spread of the virus, and make the usual linear predictions wrong, even in the short term. The outbreak of the global pandemic of the coronavirus COVID-19, which is a global medical, political, organizational, moral, economic and intellectual challenge to humanity, has stimulated not only an intensified search for drugs, but also a simulation of the mechanisms of the spread of viruses in the human population [9-12], as a better understanding mechanisms for the development of pandemics allows us to make better decisions when dealing with them. The key goal of our efforts is to obtain a decision-making tool based on the choice of the best option, within the framework of the system analysis paradigm [13], when exact decisions are impossible, but a number of characteristic scenarios can be distinguished, depending on the decisions made.

---

[1] e-mail: golovinski@bk.ru

The main factors requiring consideration in constructing a realistic model for the spread of viruses are the effects of delay, the spatial structure of a heterogeneous system, and the rate of convective transmission of infection by traffic flows. Earlier, we studied the simplest step-by-step model of the distribution of network lesions displayed in the form of a graph, and showed the critical dependence of network stability during the spread of viral-type lesions on the degree of its connectivity [14]. The analysis revealed the extreme vulnerability of global networks. As the main tool for protecting the network, dynamic clustering was proposed with isolation of unaffected network sections. In this paper, we propose a dynamic network mathematical model, as a generalization of available models and approaches, and present the result of the study of artificially constructed examples that demonstrate some possible scenarios for the development of pandemics.

## 2. The mathematical model of the spread of viruses

The simplest model of the development of viral infection in the SIR population (susceptible, infection, recovered) is described by a system of three differential equations [15]. We will extend this model to a set of interconnected clusters between which a certain proportion of infected individuals are exchanged [16] without changing the general balance of the population. This means that the flow of infected individuals is small and proportional to the total number of infected individuals in these clusters. Formally, such an assumption unreasonably violates the preservation of the total number of individuals, but we compensate for this effect by transferring uninfected individuals in the opposite direction and restoring the overall balance. The main factor for highly infectious viruses is the fact of transfer and its intensity, and not the change due to this number in subpopulations between which the transfer occurs. We denote the number of uninfected individuals in a given cluster $S_k$, the number of infected $I_k$, and the number of those who left the dynamics as a result of a cure with stable immunity or who died $R_k$. The system of balance equations describing this process has the form

$$\frac{dS_k}{dt} = -\beta_k(t)\frac{S_k I_k(t-\tau)}{N_k} - \sum_{j \neq k} \lambda_{kj}(t) I_j(t-\tau_{kj}), \quad (1)$$

$$\frac{dI_k}{dt} = \beta_k(t)\frac{S_k I_k(t-\tau)}{N_k} - g_k(t) I_k(t-T) + \sum_{j \neq k} \lambda_{kj}(t) I_j(t-\tau_{kj}),$$

$$\frac{dR_k}{dt} = g_k(t) I_k(t-T).$$

In this system of equations, the term $-g_k I_k(t-\tau)$ in the right-hand side of the first equation takes into account the elimination of infected virulent individuals from the spread of the virus due to isolation. The last term in the right-hand sides of the first two equations is responsible for the transport of viruses by traffic flows. Coefficient $g_k(t)$ represents the proportion of individuals isolated per day. In these equations $N_k = const$ is a conserved total number of individuals of all types in the $k$-th subpopulation; $\tau$ is an incubation period of the virus, causing a delay in the reproduction of the virus compared to the time of infection; $\tau_{kj}$ is a delay time associated with the transport of infected between clusters; $T$ is an average delay time from infection to isolation; $\beta_k(t)$ is an infection rate; $g_k(t)$ is a coefficient of isolation of existing patients; $\lambda_{kj}(t)$ is a coefficient characterizing the rate of transfer of infected from the $j$ - th cluster to the cluster with number $k$. Coefficients $\beta_k(t)$ can be determined by knowing the number of infected individuals per unit time. In the same way, it is possible to determine the number of people who were isolated from the number of those who became infected in a single time interval, taking into account the shift by the average time of the disease. We believe that the coefficients $g_k(t)$, $\beta_k(t)$, $\lambda_{kj}(t)$ individual for each node (city) and may depend on time due to the gradual exhaustion of resources, the introduction of an isolation regimen or a change in the quality of treatment. When analyzing scenarios, we restrict ourselves to fixed constant values of the parameters $g_k(t)$, $\beta_k(t)$, $\lambda_{kj}(t)$. We assume that the observed number of infected individuals arriving from another cluster is proportional to the number of infected individuals in the source. This means the absence of dynamics of change during the movement, i.e. short lag times typical of, for example, air travel. In this case, we can simply consider the delay time equal to zero. Otherwise, for example, when considering a vessel or a long-distance train as a vehicle, it is necessary to modify the system of equations (1), taking more fully into account the changes that occur with passengers over a long travel time. The system of equations (1) assumes that those who are ill leave, gaining immunity or dying, and are no longer at risk of infection. From previous models, our model is distinguished by the effects of delay.

### 3. Numerical examples and discussion

The system of equations (1) has a simple structure, but nonlinear, and its analytical solution is impossible, although a numerical solution is quite simple. It should also be taken into account that when deciding on the visible symptoms of the incidence, we do not see today picture (without special detection methods), but situation several days before due to the latent period of infection. In the modeling process, we assume $\tau = 5, \tau_{kj} = 0, T = 8$ days. Coefficients $\beta_k(t)$ easy to determine in the initial stages of the spread of the virus when $\beta_k(t) = \Delta I_k / I_k$, and they represents the relative increase in the number of infected in one day, if we neglect the effects of transportation between cities. If we assume that the decrease in the number of infected people in the absence of the possibility of re-infection occurs due to isolation in one form or another, then after a lapse of average time $T$ a certain proportion of previously infected people will cease to be carriers of infection.

As a first step, we show the dynamics of the spread of the virus in a particular city, described by the equations

$$\frac{dS}{dt} = -\beta \frac{SI(t-\tau)}{N}, \qquad (2)$$

$$\frac{dI}{dt} = \beta \frac{SI(t-\tau)}{N} - gI(t-T)$$

with initial conditions $I(0) = 1$, $S(0) = N$. In the absence of transfer from other subpopulations, convenient to go to the new variables $s = S/N, i = I/N$, and we get

$$\frac{ds}{dt} = -\beta si(t-\tau), \qquad (3)$$

$$\frac{di}{dt} = \beta si(t-\tau) - gi(t-T),$$

with initial conditions $i(0) = 1/N, s(0) = 1$. Thus, our equations demonstrate self-similarity, with the exception of the influence of the initial conditions. They differ from the standard SIR equations by taking into account two delay effects: the duration of the incubation period when the individual is infected but not yet a source of viruses, and the average duration from infection to isolation. In fig. 1,2,3 shows the results of numerical modeling of the solution of equations (4) for cities with a population of 10 thousand, 100 thousand, 1 million, 10 million inhabitants, without taking into account a) and b) taking into account delays at various speed indicators of infection and isolation.

In all cases, with a single individual initial infection, the effects of delay significantly worsen the picture of a pandemic. Comparison of the spontaneous dynamics of the spread of the virus and the dynamics with the restriction on contacts and isolation of infected people shows the effectiveness of quarantine measures. Thus, the best scenario for combating a pandemic is determined by the economic strength of society or its willingness to endure restrictions and hardships, especially in large metropolitan areas. It is very clear that the number of lives saved and the degradation of an idle economy are closely related. However, a decrease in the virus transfer rate alone without strict isolation measures does not produce a significant effect.

The presence of a system of cities connected by traffic streams complicates the picture of the spread of viruses [17]. We will consider several artificially constructed examples demonstrating the different dynamics of the spread of the virus between cities connected by traffic flows. As options, consider the spread of viruses in fully connected systems. We take two groups of cities: cities with a population of one million people and cities with a population of two hundred thousand people. This allows us to analyze the effect of convective transport of infection on the heterogeneous spread of viruses.

The first two equations of the system of equations (1) are solved independently of the third. Given the approximations made, we obtain a closed system of equations

$$\frac{dS_k}{dt} = -\beta_k(t)\frac{S_k I_k(t-\tau)}{N_k} - \sum_{j \neq k} \lambda_{kj}(t) I_j(t), \qquad (4)$$

$$\frac{dI_k}{dt} = \beta_k(t)\frac{S_k I_k(t-\tau)}{N_k} - g_k(t) I_k(t-T) + \sum_{j \neq k} \lambda_{kj}(t) I_j(t).$$

To further simplify the analysis, we assume that $g_k = g$, $\beta_k = \beta$, i.e., the intensities of all processes are the same across the network of cities. This extreme simplification will make it possible, by intentionally reducing the detail of the model, to understand the influence of factors of various nature outside their inevitable spread in magnitude in real systems. Примем также, что $\lambda_{kj} = \lambda = 1/1000$ for all transport links between cities. The parameters chosen by us are of an evaluative model nature and are not reliable for reproducing possible real situations. For analysis, we will separately display the percentage dynamics of the spread of viruses for large and small cities.

The results of modeling the spread of viruses in a full-connected system from one millionth city and five small cities with the first single infection in the largest of them are shown in Fig. 4: a) when $\beta = 1.4$, $g = 0.3$ и b) subject to restrictions on the spread and isolation of the population ($\beta = 1.1, g = 0.9$). The solid lines show the results for the big city, and the dashed lines show the dynamics of distribution in small cities. The results of modeling the spread of viruses in a fully connected system of five million-citizen cities and fifty small cities with the first single infection in the largest of them are shown in Fig. 5: a) when $\beta = 1.4$, $g = 0.3$ and b) under conditions of restriction of population movement and isolation of the infected population ($\beta = 1.1, g = 0.9$). The main change in the distribution dynamics for a large network is associated with the non-uniformity of the growth rate at the infection development site of large cities.

### 4. Conclusions

The effects of a incubation delay of several days lead to a doubling of the duration of the pandemic, stretching it from the more usual monthly period to the 90-100 days interval in a separate megapolis. The dynamics of infection is asymmetrical with respect to the peak shape. The increasing in infection takes longer and smoothly, compared with the completion phase of the spread of infection. Reaching a constant growth rate of infected people means that the peak of the pandemic is approaching. When viruses spread in a system of connected cities, starting with a large city, an additional delay effect arises due to the time spent on the transfer of pathogens. A feature of the dynamics of a pandemic in large cities is the occurrence of a nonmonotonic growth rate of infections and even possible local sections of the recession, which can be misleading, causing unreasonable assumptions about the near end of the pandemic and the premature removal of restrictive measures. This non-uniformity is due to the role of the environment as a reservoir for re-infection. The second wave of infection can be avoided by introducing short-term restrictions on entry into large cities during the peak phase of infection in small cities.

Our proposed modification of SIR takes into account the nonlocality in time of processes in a simplified form using delay constants, while a more detailed description can be obtained using integral operators that take into account memory effects. The development of such a more perfect model is hindered not only by theoretical difficulties, but also by the vagueness of the practical availability of determining its parameters. However, the general conclusion about the feasibility of tough measures restricting the spread of viruses with a long incubation phase, using social distance and isolation of infected people, our model shows with high visibility. We note that the simulation of real scenarios requires the use of specific data on the number of cities, transport activity, as well as the infection and isolation rate constants.

a)

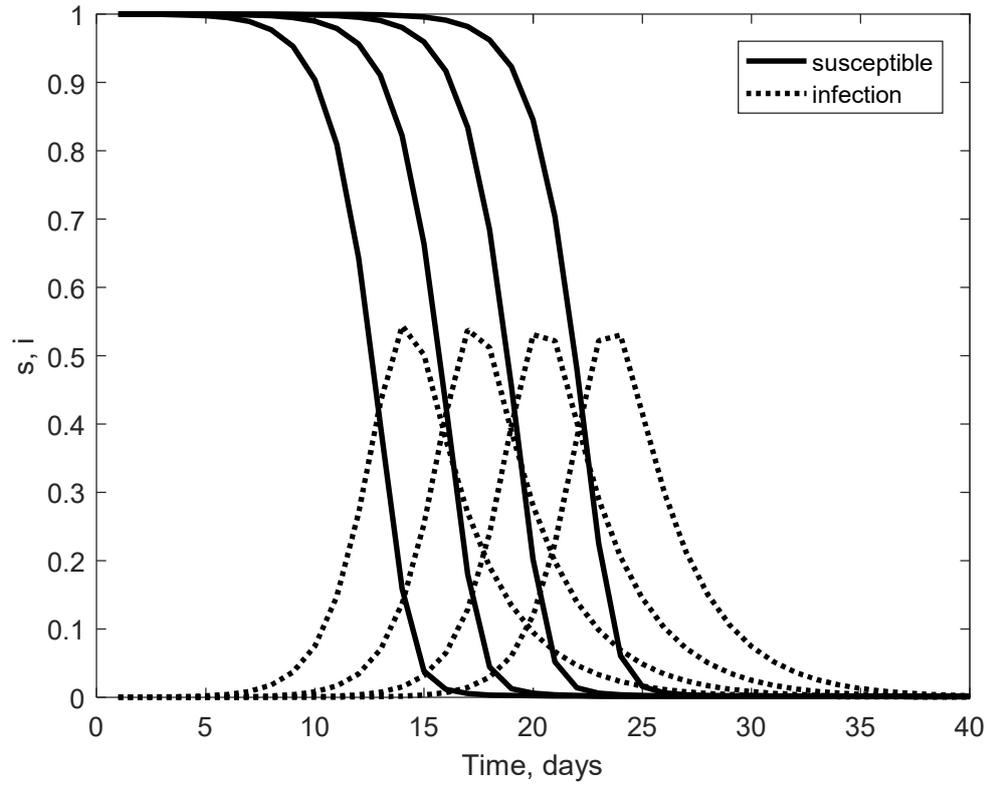

b)

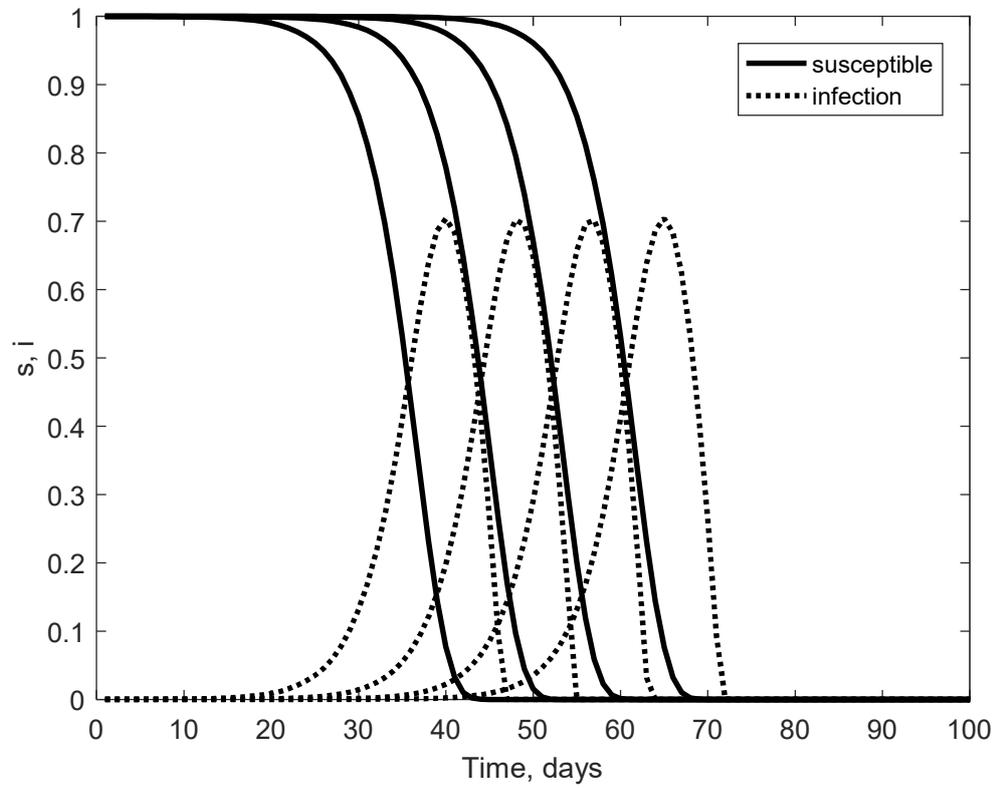

**Fig. 1.** The dynamics of the natural spread of viral infection: $\beta = 1.4, g = 0.3$.

a)

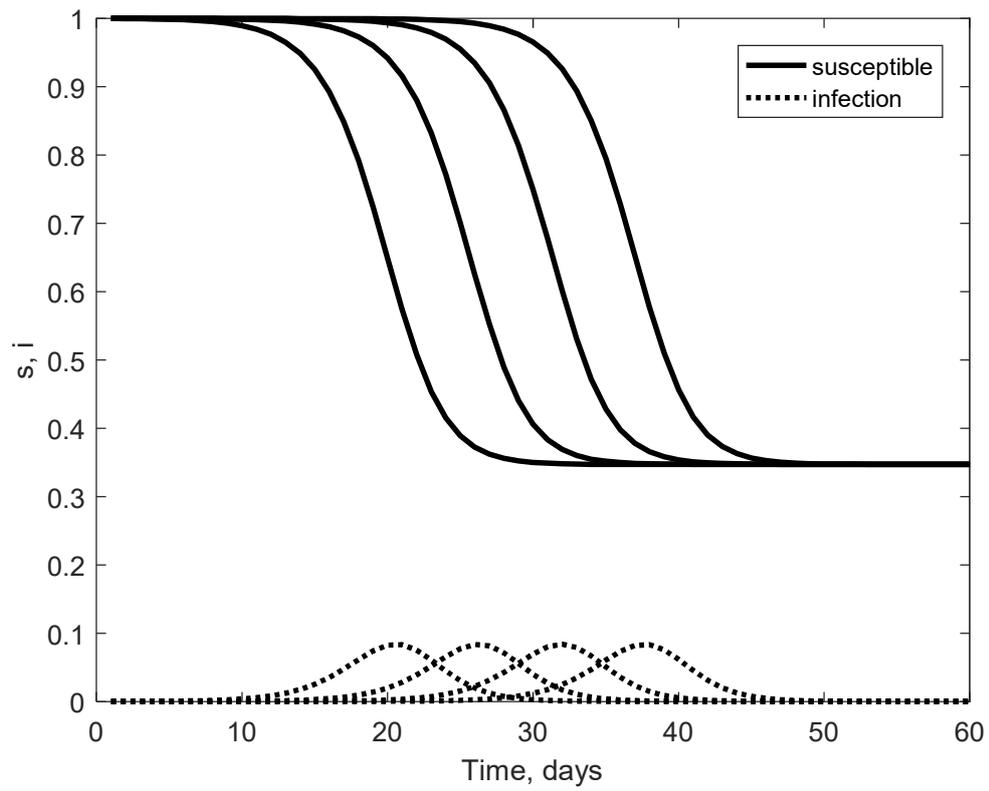

b)

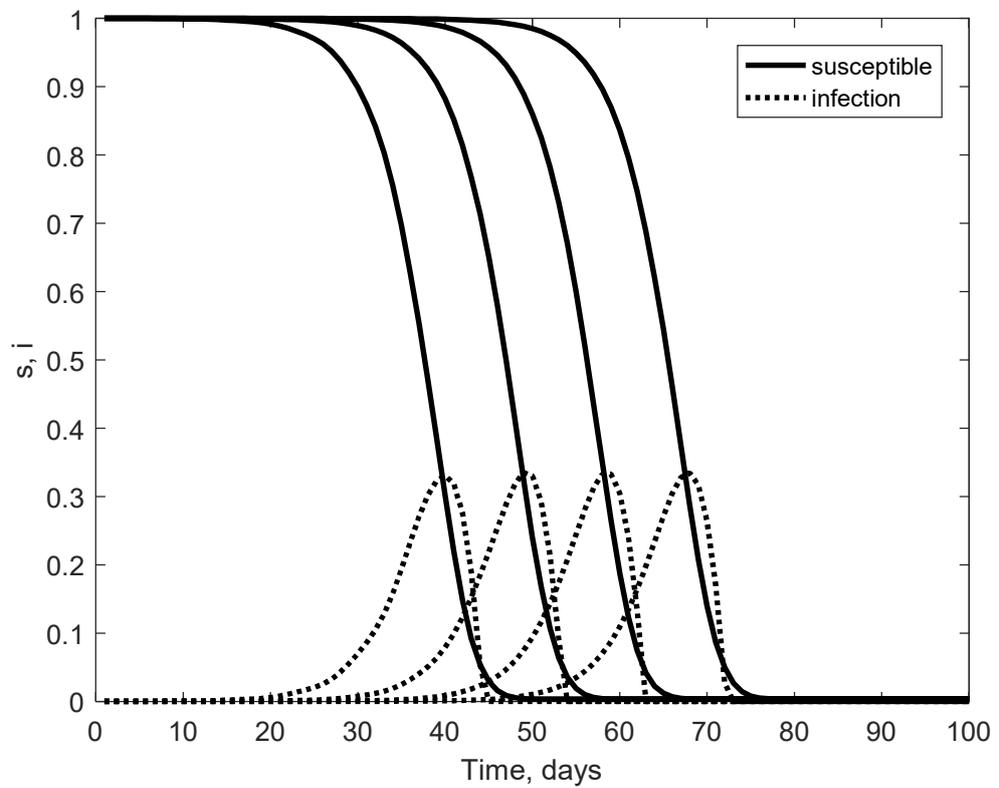

**Fig. 2.** The dynamics of the spread of viral infection with additional isolation of infected: $\beta = 1.4, g = 0.9$.

a)

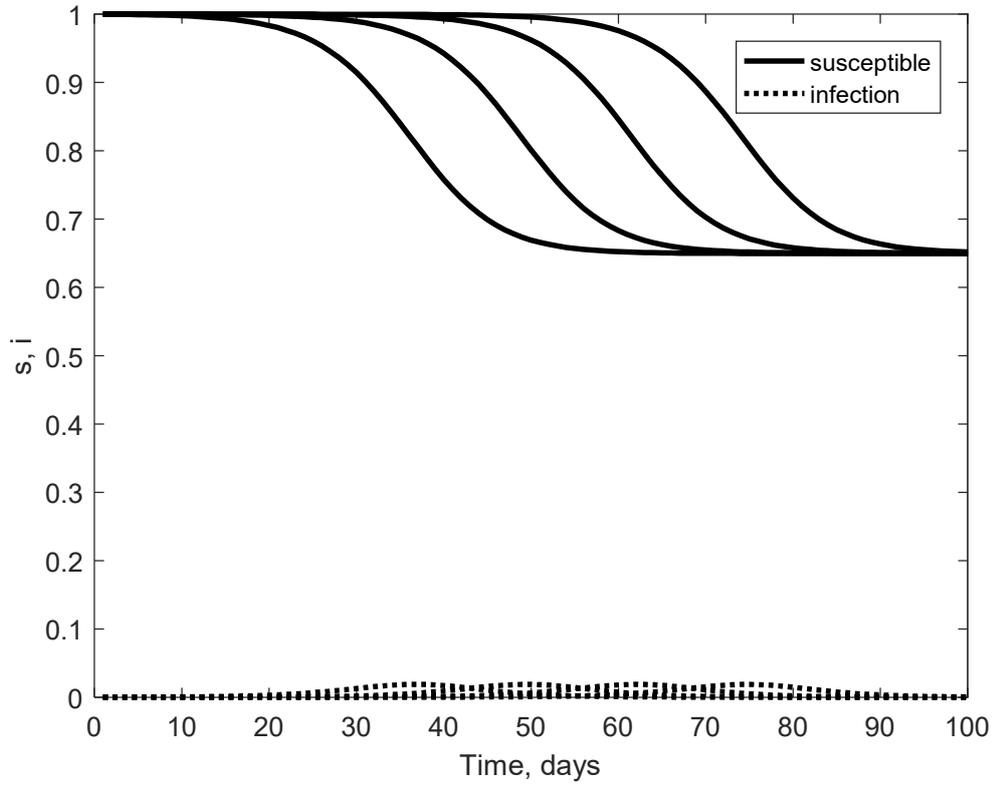

b)

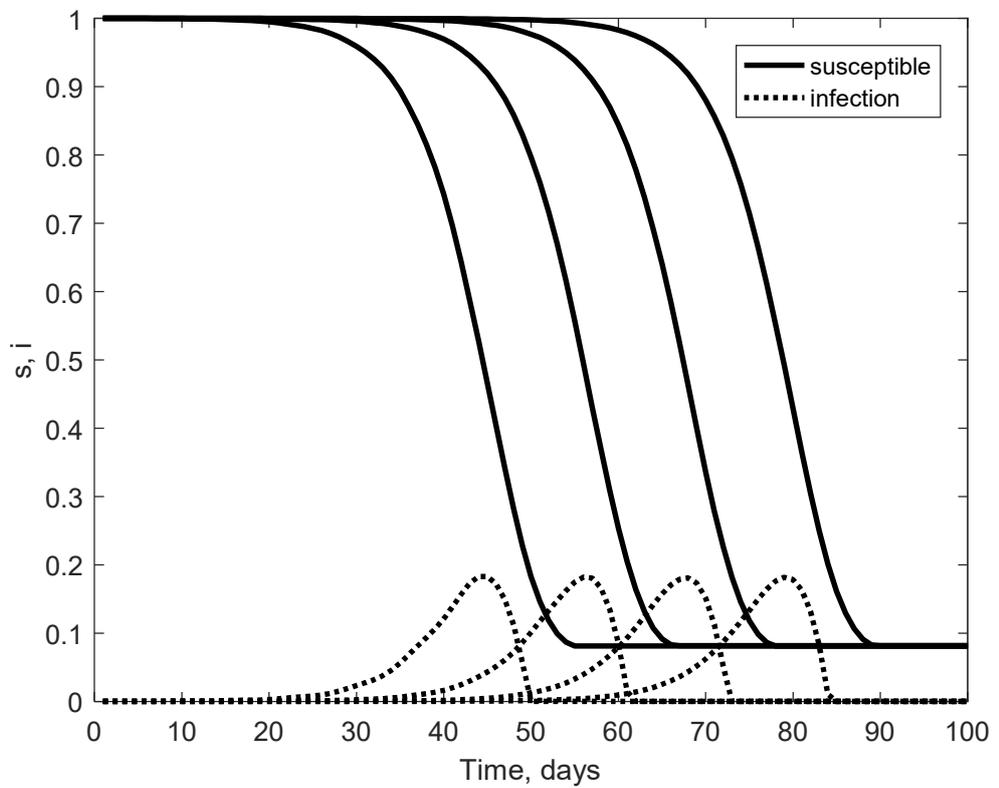

**Fig. 3.** The dynamics of the spread of viral infection with social distance and additional isolation of infected: $\beta = 1.1, g = 0.9$.

a)

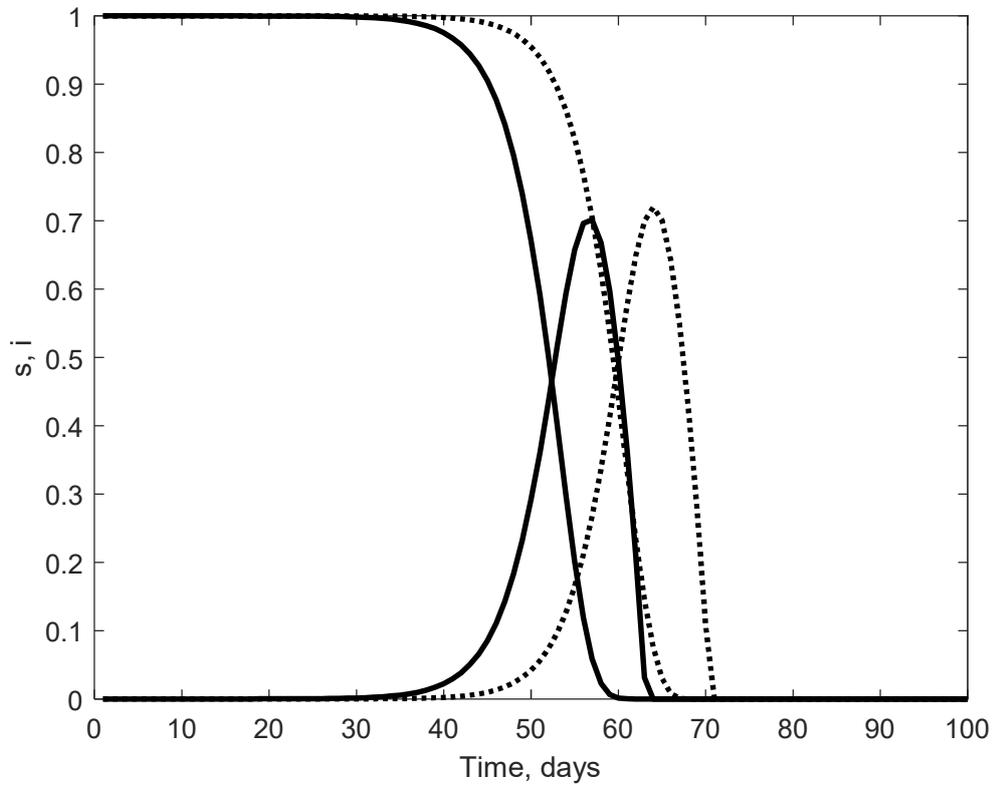

b)

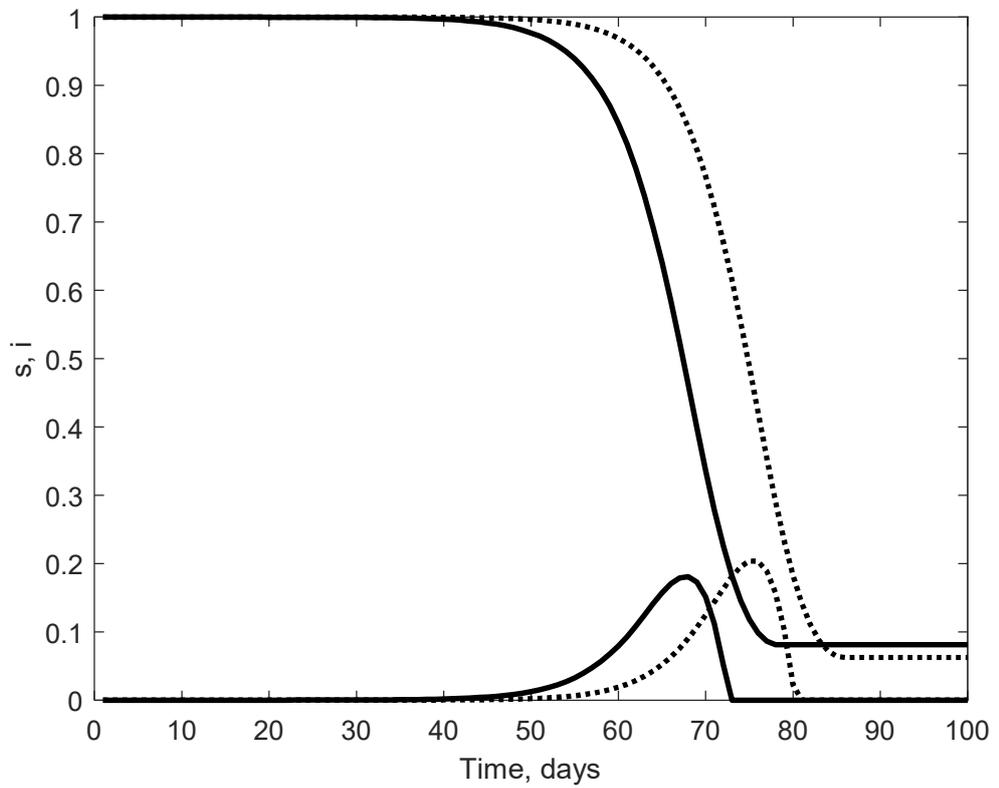

**Fig. 4.** The dynamics of the spread of the virus in a small network of connected cities without restrictions a) and with the introduction of restrictions b).

a)

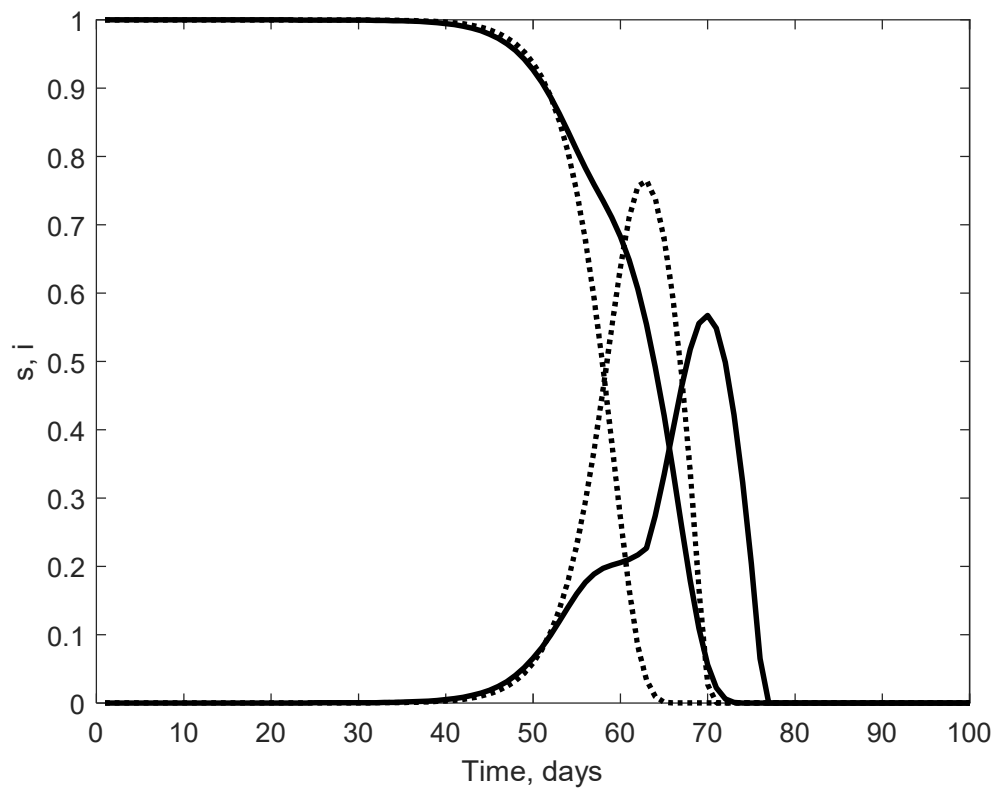

b)

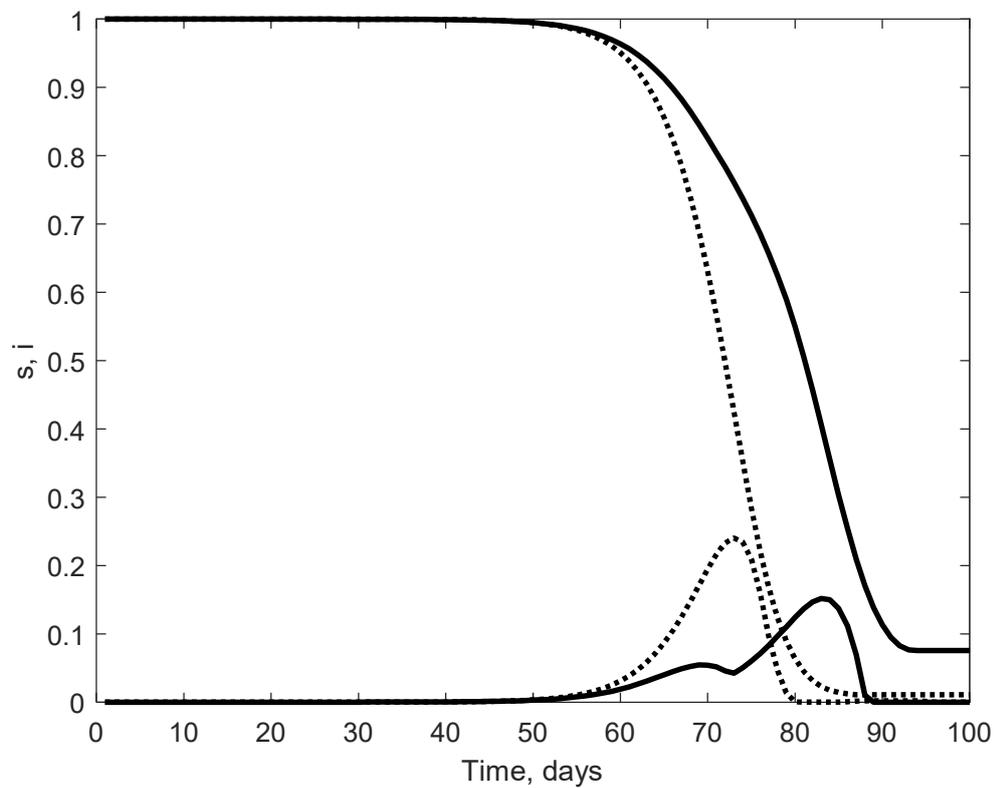

**Fig. 5.** The dynamics of the spread of the virus in a large network of connected cities: without restrictions a) and with the introduction of restrictions s b).